\documentclass[aps,prd,showpacs,nofootinbib,floats,floatfix,preprintnumbers,groupedaddress,twocolumn]{revtex4}

\usepackage{bm}
\usepackage{latexsym}
\usepackage{dcolumn}
\usepackage{amsmath,amsfonts,amssymb}
\usepackage{graphicx,epsfig}
\usepackage{amsmath}
\usepackage{fancyhdr}
\usepackage{hyperref}
\usepackage{graphicx,epstopdf}
\hypersetup{
	colorlinks   = true, 
	urlcolor     = blue, 
	linkcolor    = blue, 
	citecolor   = red 
}

{\scriptsize }

\def\p{\partial}

\begin{document}
\title{Near horizon symmetries, emergence of Goldstone modes and thermality}
\author{Mousumi Maitra\footnote {\color{blue} maitra.91@iitg.ac.in}}
\author{Debaprasad Maity\footnote {\color{blue} debu@iitg.ac.in}}
\author{Bibhas Ranjan Majhi\footnote {\color{blue} bibhas.majhi@iitg.ac.in}}

\affiliation{Department of Physics, Indian Institute of Technology Guwahati, Guwahati 781039, Assam, India
}

\date{\today}

\begin{abstract}
For a long time it is believed that black hole horizon are thermal and quantum mechanical in nature. The microscopic origin of this thermality is the main question behind our present investigation, which reveals possible importance of near horizon symmetry. It is this symmetry which is assumed to be spontaneously broken by the background spacetime, generates the associated Goldstone modes. In this paper we construct a suitable classical action for those Goldstone modes, and show that all the momentum modes experience nearly the same inverted harmonic potential, leading to an instability. Thanks to the recent conjectures on the chaos and thermal quantum system, particularly in the context of an inverted harmonic oscillator system. Going into the quantum regime, the system of large number of Goldstone modes with the aforementioned instability is shown to be inherently thermal. Interestingly the temperature of the system also turns out to be proportional to that of the well known horizon temperature. Therefore, we hope our present study can illuminate an intimate connection between the horizon symmetries and the associated Goldstone modes as a possible mechanism of the microscopic origin of the horizon thermality.        
\end{abstract}

\maketitle


\section{Introduction: BMS symmetry and  Goldstone modes}
Black hole is a fascinating object in Einstein's theory of gravity. Even though it exists for a long time, we still do not understand it fully. One of the important properties of it is its very nature as a thermal object. Over the last few years various efforts being made to understand this thermality, transpired to the fact that symmetries, their breaking and associated Goldstone modes may play fundamental role in understanding this subject. In this paper we try to establish this connection between symmetry and thermality of the black holes with some encouraging results.   

Symmetry breaking phenomena is ubiquitous in nature. Across a large span of physical problems in particle physics, cosmology and condensed matter physics, it is not only the symmetry, but it's spontaneous breaking also plays very crucial role in understanding the low energy properties. Symmetries in nature are broadly classified into two categories. A symmetry which acts globally on the physical fields are called global symmetry. Most importantly for each global continuous symmetry there exists associated conserved charge which encodes important properties of the system under consideration. Another class of symmetry which acts locally on the fields, generally known as gauge symmetry, makes the description of the system redundant. Unlike global continuous symmetries, gauge symmetry does not have associated non-trivial conserved charge \cite{AlKuwari:1990db}\cite{Karatas:1989mt}. However, as compared to the global symmetry the most striking property of a global continuous symmetry lies in its spontaneous breaking phenomena which plays very important role in understanding the low energy behaviour of the system under consideration. If a global continuous symmetry of a system breaks spontaneously, associated Goldstone boson mode emerges, whose dynamics will characterise the underlying  states and their properties of the system \cite{Weinberg:1996kr}.
On the other hand breaking of gauge symmetry is inherently inconsistent with the theory under consideration.  

In the present paper, we will be trying to understand the dynamics of the Goldstone boson modes associated with a special class of global symmetry arising at the boundary of a spacetime with nontrivial gravitational background. The generic underlying symmetry of a gravitational theory is spacetime diffeomophism which is a set of local general coordinate transformation. Therefore, diffeomrophism can be thought of as a gauge symmetry of the gravitational theory. However, it is well known that a gauge symmetry in the bulk acts as a non-trivial global symmetry at the boundary. Therefore, even if gravitational theory can be formulated as a gauge theory, theory of Goldstone modes can still be applicable and information about the microscopic gravitational states may be extracted from the boundary global symmetry. Symmetries near the boundary of a spacetime has been the subject of interest for a long time \cite{Bondi:1962px}-\cite{Barnich:2013sxa}. 

One of the popular and important examples of such a bulk-boundary correspondence is the well known global Bondi-Metzner-Sachs (BMS) group \cite{Bondi:1962px,Sachs:1962zza,Sachs:1962wk} of transformation. BMS group is an infinite dimensional global symmetry transformation which acts non-trivially on the asymptotic null boundary of an asymptotically flat spacetime. The original study \cite{Sachs:1962zza} was done on the asymptotic null boundary of an asymptotically flat  spacetime. Subsequently the analysis on another null boundary namely the event horizon of a black hole spacetime has been studied \cite{Cai:2016idg}-\cite{Akhmedov:2017ftb}. The basic idea is to find out the generators which preserve the  boundary structure of a spacetime of our interest under diffeomorphism. Usually, one encounters two types of generators, one is super-translation associated with time reparametrization and other one is super-rotation associated with angular rotation. Over the years it is observed that these generators can play a crucial role in understanding the horizon entropy of a black hole \cite{Iyer:1994ys}-\cite{Majhi:2015tpa}. Since then there is a constant effort to understand these symmetries and its role to uncover the microscopic structure of the horizon thermodynamics. Although there is no significant progress till now, but the motivation is still there which led to some of the recent attempts \cite{Strominger:1997eq}-\cite{Setare:2016qob}.

Moreover in the series of remarkable papers \cite{Weinberg:1965nx}-\cite{Ashtekar:2018lor} a deep connection between ward identities associated with the aforementioned BMS supertranslation symmetries and Weinberg's soft graviton theorem has been unraveled. It is argued that the soft photons are the Goldstone boson modes arising due to spontaneous breaking of the asymptotic symmetries. Hence an equivalence has been established between Wienberg's soft photon theorem and BMS symmetries \cite{Campiglia:2015qka}. More interestingly the same BMS transformation is shown to be closely related with the gravitational memory effect  \cite{Strominger:2013jfa}\cite{Strominger:2014pwa}. Subsequently same effect has been shown to arise near the black hole horizon as well \cite{Donnay:2018ckb}.
 
In the present paper our focus will be on the Killing horizon specifically in Rindler and Schwarzschild background. Those horizons behave like another null boundary where bulk diffeomorphism acts non-trivially in terms of BMS-{\it like} global symmetry \cite{Koga:2001vq} \cite{Iofa:2018pnf}. Associated with those global symmetry on the horizon, black hole microstates have been conjectured to be played by the soft hairs which are essentially the Goldstone boson modes associated with the symmetry broken by the macroscopic black hole state \cite{Dvali:2011aa}-\cite{Hawking:2016msc}. Although the appearance of Goldstone modes in the context of BMS symmetry exits, its dynamical behavior has not been studied in a concrete way. It is believed that the dynamics of those modes should play crucial role in understanding the microscopic nature of the black holes. Having set this motivation, in the present paper we will study the dynamics of those Goldstone modes following the standard procedure.

In order to clarify and better understand the methodology of our calculation, let us consider the emergence of Goldstone boson mode for a well known U(1) invariant complex scalar field theory with the following Lagrangiain, ${\cal L} = 1/2 (\partial_{\mu} \phi \partial^{\mu} \phi^{\dagger}) - V(\phi\phi^{\dagger})$. 
The background solution such as $\phi_0 = c$ naturally breaks U(1) symmetry which transforms the vacuum as 
\begin{eqnarray}
\phi_0' \rightarrow e^{i \pi(x)} \phi_0 = c + i c \pi(x) .
\label{R1}
\end{eqnarray}
Now we can identify the $\pi(x)$ as Goldstone boson field, and calculate the Lagrangian as follows
\begin{eqnarray}
{\cal L}_{\pi} &=& \frac 12 (\partial_{\mu} \phi_0' \partial^{\mu} \phi_0'^{\dagger}) - V(\phi_0'\phi_0'^{\dagger}) \nonumber \\ 
&=& \partial_{\mu} (c + i c \pi(x)) \partial^{\mu} (c-i  c \pi(x)) - V(c + i c \pi(x)) \nonumber \\
&=& \frac {c^2} {2} \partial_{\mu} \pi(x)) \partial^{\mu} \pi(x) + \cdots.
\end{eqnarray}
The last expression should be the leading order Goldstone boson Lagrangian associated with the broken U(1) symmetry (more detail can be found in \cite{Peskin}). Throughout our following discussions, we will use this analogy to understand the dynamics of the Goldstone mode in the gravity sector.

In the first half of our paper we consider Rindler space time with flat spatial section. In the later half we consider the asymptotically flat Schwarzchild black hole. Once we have a gravitational background, we first identify the global symmetry associated with the null boundary surface \cite{Cai:2016idg}-\cite{Akhmedov:2017ftb} \cite{Maitra:2018saa} imposing the appropriate boundary conditions. {\it Boundary conditions are such that the near horizon form of the metric remains invariant after the symmetry transformation}. However, macroscopic quantities such as mass, charge and angular momentum characterizing the physical states of a black hole under consideration will change under those symmetry transformation. Such phenomena can be understood as a spontaneous breaking of the aforementioned boundary global symmetry by the black hole background. We, therefore, expects the associated dynamical Goldstone boson modes. As mentioned earlier in this paper we will study the dynamics of those Goldstone boson modes which may shed some light on the possible microscopic states of the black holes.

\section{Rindler Background}
In this section we will consider the simplest background and try to understand the symmetry breaking phenomena as described in the introduction.
The Rindler metric, in the Gaussian null coordinate is expressed as 
 \begin{equation}
ds^2 = -2 r \alpha dv^2 + 2 dv dr + \delta_{AB} dx^A dx^B~.
\label{BRM2}
\end{equation}
The Rindler horizon is located at $r=0$. $\alpha$ is the acceleration parameter which characterizes the macroscopic state of the background spacetime. Symmetry properties of the horizon can be extracted from the following fall off and gauge conditions, 
\begin{eqnarray}
\pounds_\zeta  g_{rr}= 0, \ \ \ \pounds_\zeta  g_{vr}=0, \ \ \ \pounds_\zeta  g_{Ar}=0~;\label{con}\\
\pounds_\zeta  g_{vv} \approx \mathcal{O}(r); \ \ \ \pounds_\zeta  g_{vA} \approx \mathcal{O}(r); \ \ \  \pounds_\zeta  g_{AB} \approx \mathcal{O}(r)~.
\label{con1}
\end{eqnarray}
Here, $\pounds_\zeta$ corresponds to the Lie variation for the diffeomorphism $x^a\rightarrow x^a+\zeta^a$. The above conditions are satisfied for the following form of the diffeomorphism vector,
\begin{eqnarray}
\zeta^a \partial_a=&& F(v,y,z) \partial_v -r  \partial_v F(v,y,z)   \partial_r
\nonumber
\\ 
&&-r \partial^{A} F(v,y,z)  \partial_A~.
\label{BRM1}
\end{eqnarray}
Note that in this case we have only one diffemorphism parameter $F$ which characterizes the symmetry of the Rindler horizon. Since for constant $F$, it essentially gives the time translation, the general form of this time diffemorphsim which acts non-trivially on the $r=0$ hypersurface, is called {\it supertranslation}.  For details of this analysis, we refer to \cite{Maitra:2018saa, Donnay:2015abr, Akhmedov:2017ftb}. 

We shall see below that under the diffeomorphism (\ref{BRM1}) some of the metric coefficients will transform. This can be thought as similar to the transformation (\ref{R1}) which breaks the $U(1)$ symmetry. The corrections in the metric coefficients are determined by the supertranslation parameter $F$. Consequently the macroscopic parameters of the original metric will be modified and therefore with the analogy of $U(1)$ symmetry breaking, this can be regarded as the breaking of horizon boundary symmetry. Hence one can promote the parameter $F$ as Goldstone mode. Analogous to the U(1) Goldstone mode, here also we shall propose the underlying theory of F to be determined by the Einstein-Hilbert action. We shall find the leading order correction to this action due this aforesaid diffeomorphism which will, as for $U(1)$ symmetry breaking Goldstone, ultimately determine the dynamics of the ``Goldstone modes'' ($F$) in our present context.

In order to study the dynamics, let us first find the modified metric which are consistent with the aforementioned gauge (\ref{con}) and fall-off  (\ref{con1}) conditions. Important point to remember that the Lie variation of the metric component in our analysis is defined up to the linear order in $\zeta^a$ and hence we express the form of $\zeta^a$ (\ref{BRM1}) valid up to linear order in $F$.  Under this diffeomorphism vector (\ref{BRM1}), the modified metric takes the following form:  
\begin{eqnarray}
g'_{ab} &=& \Big [ g^{(0)}_{ab} + \pounds_\zeta g^0_{ab} \Big] dx^a dx^b  \nonumber\\
&=& -2 r \alpha dv^2 + 2 dv dr + \delta_{AB} dx^A dx^B \nonumber\\
 &+& \Big[-2r\Big(\alpha \partial_{v} F + \partial^2_v F\Big)\Big] dv^2 \nonumber\\
 &+& \Big[-2r\Big(\alpha \ \partial_{A}F + \partial_{A} \partial_v F\Big)\Big] dv dx^A\nonumber\\
 &+&\big[- 2 r \partial_A \partial_B F \Big] dx^A dx^B~.
\label{rindler}
\end{eqnarray}
In the above, $g^{(0)}_{ab}$ is the original unperturbed metric (\ref{BRM2}), whereas all linear in $F$ terms are incorporated in $h_{ab}$. Under the following supertranslation symmetry transformation,
\begin{equation}
v'= v + F(v,x^A)~,~x'^A = x^A - r\partial^A F(v,x^A) ,
\end{equation}
we can clearly see the macroscopic state parameter $\alpha$ of the original Rindler background transforms into
\begin{equation}
\alpha \rightarrow  \alpha  +\Big(\alpha \partial_{v} F + \partial^2_v F\Big)\label{alpha} .
\end{equation}
Therefore, this change of macroscopic state by the symmetry transformation can be understood as a breaking of the boundary symmetry of the Rindler spacetime \cite{Eling:2016xlx}. As $F$ is the parameter associated with the broken symmetry generator, following the standard procedure of Goldstone mode analysis, we promote $F$ as a Goldstone boson field. However, all the measure will be done with respect to the usual unprimed coordinate, and dynamics of the mode is defined on the $r=0$ hyper-surface.   

Since $\alpha$ appears as a Lagrange multiplier in the Hamiltonian formulation, one usually chooses the gauge where variation of alpha is zero everywhere \cite{Eling:2016xlx,Eling:2016qvx}. However strictly speaking this is not a generic choice. It is sufficient to set the variation of $\alpha$ to be zero only at the boundary for consistency, 
\begin{equation}
 \delta \alpha (-\infty,x^A) = \lim_{v\rightarrow-\infty} \Big(\alpha \partial_{v} F + \partial^2_v F\Big) =0, \label{boundary}
\end{equation}
where horizon is located at $v\rightarrow -\infty$.
One of the obvious choices to satisfy the above condition is to set the total variation $\delta \alpha$ to be zero everywhere \cite{Donnay:2016ejv}. This naturally set the boundary condition at the horizon and furthermore makes the field $F$ non-dynamical. Therefore, we believe this restrictive condition does not capture the full potential of the Goldstone modes. Our goal of this paper is to go beyond and understand the dynamics of this Goldstone modes, which could be the potential candidate for the underlying degrees of freedom of the black hole. Therefore, 
we first construct an appropriate Lagrangian of this mode and finally at the solution level we set the boundary condition such that Eq. (\ref{boundary}) is automatically satisfied at the horizon. Important to note that if one allows the fluctuation of $\alpha$ even at the boundary, one needs to take care of the appropriate boundary terms (e.g. see \cite{Bunster:2014mua}\cite{Perez:2016vqo}), which will be discussed in a separate paper.

\subsection{Dynamical equation for $F$} 
As we have already pointed out, in order to study the dynamics for $F$ we propose the  Lagrangian $\mathcal{L}_{F}$ associated with the newly perturbed metric (\ref{rindler}) near the $r=0$ surface: 
\begin{eqnarray}
\mathcal{L}_{F} =  \sqrt{-g'} R' \label{L}~.
\end{eqnarray}  
Here $R'$ is the Ricci scalar calculated for the newly constructed metric $g'_{ab}$ (\ref{rindler}) and $g'$ is the corresponding determinant. To study  the dynamics of the Goldstone mode associated with the horizon symmetry, first we will compute the above Lagrangian eq. (\ref{L}) at arbitrary $r$ value in the bulk spacetime and then we take the limit $r\rightarrow 0$. This procedure is similar to the stretched horizon approached in black hole thermodynamics (for example see the discussion in section $4$ of \cite{Carlip:1999cy}). In this approach, if we are interested to find any quantity on a particular surface (say $r=0$), one usually first calculate the same just away from this surface (say $r=\epsilon$, with $\epsilon$ is very small). After that the obtained value is derived by taking the limit $\epsilon\rightarrow 0$.


 Now we are in a position to expand our Lagrangian (\ref{L}) in terms of the transformed metric (\ref{rindler}). If the background metric components are  $g_{ab}=g_{ab}^{(0)}+h_{ab}$, with $h_{ab}$ as a small fluctuation, in general the Taylor series expansion of the Lagrangian around background metric $g_{ab}=g_{ab}^{(0)}$ can be written as
\begin{equation}
\mathcal{L_{F}} = \mathcal{L_{F}}(g_{ab}^{(0)}) + h_{ab}\Big(\frac{\delta\mathcal{L_{F}}}{\delta g_{ab}}\Big)^{(0)}  + h_{ab}h_{cd}\Big(\frac{\delta^2\mathcal{L_{F}}}{\delta g_{ab} \delta g_{cd}}\Big)^{(0)}+\dots
\label{BRM4}
\end{equation}
The first term of the above equation obviously does not contribute to the dynamics. Given the background metric to be a solution of equation of motion, the second term vanishes as it is essentially proportional to the Einstein's equation of motion. Third term introduces the quadratic form for the Goldstone field $F$. For our purpose of the present paper, we will restrict only to Lagrangian for the Goldstone mode which is at the quadratic order. All the higher order in $F$-terms we left for our future discussions.


The final form of the Lagrangian (\ref{L}) after taking the near Horizon limit comes out as:
\begin{eqnarray}
{\mathcal{L}_{F}} &=& \lim_{r \to 0} \Big(\sqrt{-g'} R'\Big) 
\nonumber\\
&=&  \Big[-6 \alpha^2 \partial_y F \partial_y F -6 \alpha^2 \partial_z F \partial_z F 
+4 \alpha \partial_v F \partial^2_y F \nonumber\\ & -& 12 \alpha \partial_z F \partial_v \partial_z F - 6(\partial_v \partial_z F)^2 - 12 \alpha \partial_y F \partial_v \partial_y F \nonumber\\ &-& 6(\partial_v \partial_y F)^2 + 4 \partial^2_y F \partial^2_v F + 4 \partial^2_z F (\alpha \partial_v F + \partial^2_v F) \Big]~.\nonumber\\  
\label{RL}
\end{eqnarray}
Since ${\mathcal{L}_{F}}$ is calculated on a $r=$ constant surface, the action can be defined as the integration of the above Lagrangian on $v,y$ and $z$. Th induced horizon geometry has flat spacial section, we, therefore, consider the following generic form of  $F$:
\begin{eqnarray}
F_{mn} = f_{mn}(v) \frac{1}{\alpha} \exp\Big[i (my+nz)\Big]~.
\label{F}
\end{eqnarray}
Hence the general solution for Goldstone mode would be,
\begin{eqnarray}
F(v,y,z) = \sum_{m,n} C_{mn} F_{mn}~. 
\end{eqnarray}
 Here we need to find the form of $f_{mn}(v)$ from the solution of the equation of motion obtained from (\ref{RL}). It is quite obvious that substitution of the above ansatz and then integrating over transverse coordinates one can get an one dimensional action which determine the evaluation of $f_{mn}(v)$ with respect to $v$.  Since the total derivative terms in action keep the dynamics unchanged, it may be verified that under the integration of transverse coordinates third, fourth, sixth and ninth terms are total derivative terms with respect to $v$.  So, those terms can be neglected.
Ignoring total derivative terms the final form of the Lagrangian (\ref{RL}) is given  as;
\begin{eqnarray}
{\mathcal{L}_{F}} &=&  \Big[-6 \alpha^2 \partial_y F \partial_y F -6 \alpha^2 \partial_z F \partial_z F- 6(\partial_v \partial_y F)^2 \nonumber\\ &-& 6(\partial_v \partial_z F)^2+ 4 \partial^2_y F \partial^2_v F + 4 \partial^2_z F  \partial^2_v F \Big]~.  \label{Lfi}
\end{eqnarray}

Before proceeding further, here we want to mention an important point of our proposed form of the Goldstone boson Lagrangian. Since the modified metric (\ref{rindler}) has been constructed by taking into account a particular type of diffeomorphism, one always concludes that the Lagrangian must be invariant upto a total derivative. Contribution of the total derivative term vanishes over the closed boundaries which encloses the bulk of the manifold. For instance, in the case of $\sqrt{-g}R$, the variation of it under diffeomorphism $x^a\rightarrow x^a+\xi^a$ leads to $\sqrt{-g}\nabla_a(R\xi^a)=\partial_a(\sqrt{-g}R\xi^a)$, which is a total derivative term. 
In this analysis we are interested to build a theory on the horizon (i.e. $r=0$) and horizon is a part of the closed boundary of the bulk manifold. Therefore it is expected that the total derivative term will give  non-vanishing contribution on a part of the closed surfaces such as horizon. In the case of $\sqrt{-g}R$, the
 boundary term on $r=$ constant surface in the variation of action is given by $\int d^3x\hat{n}_a\xi^a \sqrt{-g}R$, where $\hat{n}_a$ is the normal to the surface with components $(0,1,0,0)$. Therefore, on this surface our proposal for the Lagrangian density (loosely call it as Lagrangian) is $\sqrt{-g}R$.  This is precisely considered here. The Lagrangian (\ref{RL}) is not the one which is defined for the whole
  spacetime, rather it is calculated on the $r=$ constant surface, and hence coming out to be non-trivial. In this sense our proposed
   Lagrangian does not carry any ambiguity and correctly describe the dynamics of the Goldstone mode $F$ associated with the super-transnational symmetry near the horizon.

Next we concentrate on Gibbons-Hawking-York (GHY) boundary term 
\begin{equation}
\mathcal{S}_2 =- \frac{1}{8\pi G}\int d^3x \sqrt{h}K~,
\label{GHY}
\end{equation}
which is usually added to the action in order to define a proper variation of the action. The trace of the extrinsic curvature of the boundary surface ($r\rightarrow 0$) is given by $K = -\nabla_a N^a$, where $N^a$ is considered as the unit normal to the $r=constant$ hyper-surface. For metric (\ref{rindler}), its lower component is given by $ N_a =(0, 1/\sqrt{2r(\alpha +\alpha \partial_v F +\partial^2_{v} F)},0,0) $. Therefore in the near horizon limit ($r \rightarrow 0$), one gets the following form of the action coming from the GHY term:
\begin{eqnarray}
&&\mathcal{S}_2= -\frac{1}{8 \pi G}\int d^3 x\Big[\alpha +\Big(\alpha \p_v F + \frac{1}{2}\p^2_v F + \frac{1}{2 \alpha} \p^3_v F\Big)
\nonumber\\ 
&&+ \frac{1}{2 \alpha
^2}\Big(\alpha^2 \p_v F \p^2_v F + \alpha (\p^2_v F)^2 + \alpha \p_v F \p^3_v F 
\nonumber
\\
&&+ \p^2_v F \p^3_v F \Big)\Big]~.
\label{Kac}
\end{eqnarray}
However, we observed that this terms does not contribute to our required equation of motion.
In fact, the above boundary term in the action is turned out to be related to the horizon entropy which is discussed in appendix (\ref{App2}).

Note that the aforesaid  Lagrangian (\ref{RL}) contains higher derivative terms of $F$. Therefore, the theory of Goldstone boson modes emerging on the boundary of a gravitational theory turns out to be higher derivative in nature. However, if we want to trace back the origin of this higher derivative action, it is from the diffeomorphically transformed metric components which already contains the derivative term. However, we will see those higher derivative terms will be crucial for our subsequent discussions on the horizon properties. This connection could be an interesting topic to investigate further. However, one of the important point here is that at the background label system is not Lorentz invariant. The generalized Euler-Lagrangian equation, defined for higher derivative theory, is:
\begin{eqnarray}
\frac{\partial L}{\partial F} - \partial_{\mu}(\frac{\partial L}{\partial (\partial_{\mu} F)}) + \partial_{\mu} \partial_{\nu} (\frac{\partial L}{\partial (\partial_{\mu} \partial_{\nu} F)}) =0 .\label{GenEH}
\end{eqnarray}
With this the equation of motion is found to be
\begin{eqnarray}
3 \alpha^2 \partial^2_y F +3 \alpha^2 \partial^2_z F -4 \partial^2_y \partial^2_v F -4 \partial^2_z \partial^2_v F =0~.
\label{BRM5}
\end{eqnarray}
Important to note again, the contribution on the equation of motion comes only from from (\ref{Lfi}). GHY (\ref{Kac}) term does not contribute.
Substitution of (\ref{F}) in (\ref{BRM5}) yields
\begin{eqnarray} \label{modeeq1}
(m^2 +n^2) [\partial^2_{v}f_{mn}(v) -\frac{3 \alpha^2}{4} f_{mn}(v)]= 0~.
\end{eqnarray}
Important point to note that every individual mode ($m,n$), will follow the same equation of motion of a simple oscillator in an inverted harmonic potential. The solution will be,
\begin{eqnarray}
f_{mn}(v) && = A \exp\Big[(\sqrt{3/4}) \alpha v\Big] + B \exp\Big[-(\sqrt{3/4}) \alpha v \Big] \nonumber\\ &+ & f_1 (y,z) \delta_{m,0} \delta_{n,0}\label{Gsol}~,
\end{eqnarray}
for all $m,n$.
In the above, $A$ and $B$ are arbitrary constants to be determined. So far we talked about the classical dynamics of the Goldstone mode. It is apparent that the system is unstable because of the inverted harmonic potential at least at the tree level Lagrangian.This is also apparent from the solution (\ref{Gsol}). As we are interested to the near horizon region where $v\rightarrow -\infty$, the above solution grows rapidly and makes the mode unstable. Therefore, the appropriate boundary condition one can set is $B=0$, leading to
\begin{eqnarray}
F_{mn} (v,y,z) &&=  [A \exp[(\sqrt{3}/2) \alpha v]  \nonumber\\ &+& f_1 (y,z) \delta_{m,0} \delta_{n,0} ] \ \frac{1}{\alpha }\exp [i (my+nz)]~.\nonumber\\ \label{gsol2}
\end{eqnarray}
Interesting this is precisely the boundary condition which satisfies the condition of vanishing fluctuation of surface gravity $\delta \alpha=0$ at the horizon defined by the Eq. (\ref{boundary}).
 
We already know that the horizon is a special place in the entire spacetime region, as any two hypothetical observers spatially separated by the horizon can never communicated to each other. Therefore, it would have been unusual, had there been just simple stable free field like Lagrangian for the Goldstone modes. The connection between these special nature of the horizon and emergence of instability is the subject of study for a long time. Our goal of this paper would be to shed some light on this issue. {\em Does the emergence of the inverted harmonic potential has anything to do with the  thermal nature of the black hole horizon?} Of course in order to understand this, we need to go to beyond the classical regime. In the next section we will try to make this connection considering a recent proposal \cite{Morita:2018sen}\cite{Morita:2019bfr}.
\subsection{Thermal behaviour of the field solution} \label{thermal}
In this section we consider the quantum mechanical treatment of the Goldstone boson mode discussed so far. 
It has recently been conjectured that Lyapunov exponent $\lambda$ of a thermal quantum system, in presence of quantum chaos, is bounded by the temperature $T$ of the system as $\lambda \leq 2\pi T/\hbar$ \cite{Maldacena:2015waa}. Based on this result further conjecture has been made in the reference \cite{Morita:2019bfr}\cite{Kurchan:2016nju} which says that a chaotic system with a definite Lyapunov exponent could be fundamentally thermal by reversing the above inequality. To justify the argument, one of the interesting example the author has studied is the semi-classical dynamics of a particle in an inverted harmonic potential{\footnote{The choice of the inverted harmonic oscillator stems from the fact that the particle motion is unstable under this potential and hence, at the classical level, any small perturbation can lead induction of chaos in the motion (for example, see \cite{Bombelli:1991eg}).}}, and showed that the quantum correction induces an energy emission by the particle under study obeying thermal probability distribution. 
Therefore, the connection between the semi-classical chaotic system and the thermal nature is emerged.  
Interestingly, for our present system, each individual Goldstone boson mode behaves like an inverted harmonic oscillator. Hence, the aforesaid connection between the thermal emission and the semi-classical chaotic dynamics could be a potential reason for the thermal nature of the black hole horizon. Even more interestingly, every individual Goldstone boson mode parametrized by $(m,n)$ see the same inverted potential, which may also indicate the universality of the thermal nature of the horizon.
{\em Our present claim is ambitious and exciting which needs detailed future exploration}. 
Before we resort to our discussion of thermal nature of the black hole, let us briefly describe the connection between the thermality and the inverted harmonic oscillator, following from the reference \cite{Morita:2018sen}\cite{Morita:2019bfr}.  
These are connected with the finite quantum mechanical transition probability through a potential barrier. The equation of motion of the particle moving in a harmonic potential is given by,
\begin{eqnarray}
\mu \ddot{x} -\omega x =0 \label{1dParticle}
\end{eqnarray}
Here potential $V= -\frac{\omega x^2}{2}$, and $\mu$ is the mass of the particle. Important case would be, if one considers the energy of the particle $E < 0$, 
for which the potential energy of the particle is greater than its kinetic energy. With this energy if the particle travels toward the potential from the left ($x<0$), classically it cannot pass through the potential towards right $(x >0$). However, quantum mechanically the particle will have finite tunneling probability to go across the potential barrier. Therefore, the particle will have finite probability of transmission through the barrier even for $E<0$. In the similar manner for $E>0$, the particle will have finite quantum mechanical probability of reflection off the barrier, which otherwise was not possible classically.
 
Therefore, to describe the above quantum mechanical phenomena the appropriate  Hamiltonian for the wave function $\Phi(x)$ associated with the particle is expressed as 
 \begin{eqnarray}
H = -\frac{{\hbar}^2}{2}\frac{d^2}{d x^2} - \frac{{\omega} x^2}{2}
 \end{eqnarray}
with the Schr\"{o}dinger equation,
 \begin{eqnarray}
 -\frac{{\hbar}^2}{2}\frac{d^2 \Phi}{d x^2} - \frac{{\omega} x^2}{2} \Phi =E \Phi~. 
 \end{eqnarray}
$E$ is the energy of the paticle. The well known expression for the probability of transmission($P_T$) and the reflection ($P_R$) using WKB approximation (detail in \cite{Barton:1984ey})  are given as, 
\begin{eqnarray}
P_{T/R} = \frac{1}{e^{\frac{2 \pi}{\hbar} \sqrt{\frac{\mu}{ \omega}} |E|} +1} = \frac{1}{ e^{\beta |E|} +1}~. 
\label{trans}
\end{eqnarray}
An interesting interpretation of this expression is that for large absolute value of the energy $E$, probability amplitude from classical path to quantum transmission or reflection will be $\exp[-\beta |E|]$.
Therefore, the quantum harmonic oscillator system can be mapped to a
two level system with temperature $T$, whose ground state is represented as the classical trajectories and excited state is quantum one. And the temperature of the system can be easily identified as 
\begin{eqnarray}
T = \frac{\hbar}{2 \pi} \sqrt{\frac{\omega}{\mu}} ~.
\label{R2}
\end{eqnarray}
For further detail of this interesting interpretation the reader can look into the reference \cite{Morita:2018sen} \cite{Morita:2019bfr}). In this context it is worth to mention that recently one of the authors of this paper also showed by an independent and completely different way that the inverted harmonic oscillator gives rise to temperature at the quantum level \cite{Dalui:2019esx}.

In our present analysis we have obtained the dynamical equation of motion for individual mode as given in (\ref{modeeq1}). Comparing this with Eq. (\ref{1dParticle}) one can easily conclude that the dynamics of the mode along $v$ is governed by inverted harmonic oscillator potential. To clarify our analogy, each mode $f_{mn}(v)$ can be thought of as the position $x(t)$ of a particle of mass unity with $v$  playing the role of time coordinate as $t$. Therefore, we have following equivalence table:
\begin{eqnarray}
 f_{mn} \equiv x; \,\,\,\,\ v \equiv t; 
\end{eqnarray}
accompanied by the identifications
\begin{equation}
\mu \equiv 1;\,\,\,\,\
\omega \equiv \frac{3 \alpha^2}{4}~.
\end{equation}
%
Hence by the earlier argument, we can conclude that each mode, at the quantum level, is thermal. The temperature is evaluated as (\ref{R2}) with the following substitutions: $\mu=1$ and $\omega=(3\alpha^2/4)$. Therefore in our case it is given by 
\begin{eqnarray}
T = \frac{\hbar}{2 \pi} \frac{\sqrt{3}\alpha}{2} ~.
\end{eqnarray}

Even more interestingly what is emerged from our present calculation that all the modes with quantum number $(p,q)$ are degenerate with respect to $E$. {\em This observation seems to suggest that the horizon under study can carry entropy because of those degenerate quantum states}. However, in order to have finite entropy, we need to have an upper limit on the value of $(p,q)$, which must be proportional to the only scale available in the theory namely Planck scale.
%
%
Our naive analysis based on \cite{Morita:2018sen}, shows that semi-classical Goldstone boson dynamics can capture the well known thermal behaviour of the horizon. {\em Moreover, the temperature turned out to be proportional to the acceleration of the Rindler frame}. This is an important observation as we know that the Rindler horizon is thermal with respect to its own frame. In this case also the temperature is proportional to $\alpha$, known as {\it Unruh temperature} \cite{Unruh:1976db}. 
However, the proportionality constant appeared to be different. 	
 {\em Another important outcome of our analysis is the emergence of infinite number of degenerate states which can be associated with the entropy on this horizon}. We will take up this issue in our future publication. The microscopic origin of the horizon thermodynamics is a subject of intensive research for a long time. Our present analysis hints towards an important fact that the BMS like symmetry near the horizon could play important role in understanding the thermal nature and possible origin of the underlying microscopic states of a black hole. Motivated by our analysis, in the subsequent section we will discuss about the Schwarzschild black hole. 

\section{Schwarzschild black hole}
So far we have discussed about the dynamics of Goldstone boson mode in Rindler background. To this end we perform similar analysis considering Schwarzschild black hole background. 
The near horizon geometry of the Schwarzchild black is again Rindler, however, with two dimensional sphere at each point. Therefore, we expect similar behavior of the Goldstone mode for this case as well. As we go along we also notice the main differences with flat Rindler case. 

The Schwarzschild metric in Eddington-Finkelstein coordinate ($v,r,\theta,\phi$) is expressed as,
 \begin{equation}
ds^2 = -(1- 2M/r) dv^2 + 2 dv dr + r^2 \gamma_{AB} dx^A dx^B~.
\label{metric}
\end{equation}
The event horizon is located at $r=2 M$. $M$ is the mass of the black hole  which characterizes the macroscopic state of the background spacetime. Asymptotic symmetry properties of the horizon can be extracted from similar fall off and gauge conditions for the metric components,  
\begin{eqnarray}
\pounds_\zeta  g_{rr}= 0, \ \ \ \pounds_\zeta  g_{vr}=0, \ \ \ \pounds_\zeta  g_{Ar}=0~;\label{con2}\\
\pounds_\zeta  g_{vv} \approx \mathcal{O}(r); \ \ \ \pounds_\zeta  g_{vA} \approx \mathcal{O}(r); \ \ \  \pounds_\zeta  g_{AB} \approx \mathcal{O}(r)~.
\label{con3}
\end{eqnarray}
Here, $\pounds_\zeta$ corresponds to the Lie variation for the diffeomorphism $x^a\rightarrow x^a+\zeta^a$. 
The primary motivation to consider the aforementioned conditions is essentially to preserve the form of the metric under the diffeomorphism. As has already been observed in our previous case, those differmorphsim in turn renormalizes the state of the black hole parameter such as mass $M$ of the Schwarzschild black hole. Similar to our previous analysis after solving the above gauge fixing conditions with the imposed fall-off conditions, the diffeomorphism vectors turned out to be,
\begin{eqnarray}
\zeta^a \partial_a= F(v,x^A) \partial_v -(r-2M)  \partial_v F   \partial_r \nonumber\\ + (1/r -1/2M) \gamma^{AB} \partial_{B} F \ \partial_A .
\end{eqnarray}
Again we have one unknown function $F$ which is identified as  supertranslation generator. Under this transformation the background metric takes of following form \cite{Averin:2016ybl},
\begin{eqnarray}
g'_{ab} &=& \Big [g^{(0)}_{ab} + \pounds_\zeta g^0_{ab} \Big] dx^a dx^b\nonumber\\
 &=& -(1- 2M/r) dv^2 + 2 dv dr + r^2 \gamma_{AB} dx^A dx^B \nonumber\\
 &+& \Big[2M/r(1- 2M/r) \partial_v F - 2 (1- 2M/r)\partial_v F \nonumber\\&& -2 (r-2M) \partial^2_v F \Big] dv^2 + \Big[-(1-2M/r) \partial_{A} F \nonumber\\ &&-(r-2M) \partial_{A}\partial_v F + r^2 \partial_{A}\partial_v F (1/r - 1/2M) \Big] dv dx^A \nonumber\\ &+& \Big [-2 (2M-r)r \gamma_{AB} \partial_v F\nonumber\\ &-& (1/r -1/2M)(\partial_E F \gamma^{DE} \partial_D \gamma_{AB}\nonumber\\ &+& \gamma_{AD} \ \partial_B (\partial_E F \gamma^{DE})) \Big] dx^A dx^B~.
\label{SS}
\end{eqnarray}
As has already been discussed for the Rindler metric with flat spatial section, for the present case the modification $h_{ab}$ due to following super-translation,
\begin{equation}
v'= v + F~;~x'^A = x^A +  (1/r -1/2M) \gamma^{AB} \partial_{B} F ,
\end{equation}
the macroscopic black hole state parameter $M$ renormalizes to,
\begin{equation}
\frac{1}{M} \rightarrow  \frac{1}{M} +\frac{1}{M}\Big(\partial_{v} F + 4 M \partial^2_v F\Big) .
\end{equation}
Therefore, this change of macroscopic state by the symmetry transformation can similarly be understood as a breaking of the boundary super-translation symmetry with $F$ as the broken symmetry generator. 

Following the same procedure as for the Rindler case, the Lagrangian  $\mathcal{L}_{F}$ of the Goldstone mode on the horizon surface takes the following form,
\begin{eqnarray}
\mathcal{L_{F}}&=& \Big[  \frac{-3}{2 (2M)^2} \csc\theta \ \partial_{\phi} F \partial_{\phi} F - \frac{3}{2 (2M)^2} \sin\theta \ \partial_{\theta} F \partial_{\theta} F \nonumber\\ &+& 4 \sin\theta \ \partial_{v} F \partial_{v} F - \frac{3}{M} \csc\theta \ \partial_{\phi} F \partial_{v} \partial_{\phi} F \nonumber\\ &+& \frac{1}{M} \cos\theta \ \partial_{\theta} F \partial_{v} F -\frac{3}{M} \sin\theta \ \partial_{\theta} F \partial_{\theta} \partial_{v} F \nonumber\\ &+& 4 \cos\theta \ \partial_{\theta} F \partial^2_{v} F + \frac{1}{M} \csc \theta \ \partial^2_{\phi} F \partial_v F\nonumber\\ & +& 4 \csc\theta \ \partial^2_{\phi} F \partial^2_{v} F + \frac{1}{M} \sin\theta \ \partial^2_{\theta} F \partial_v F\nonumber\\ & +& 4 \sin\theta \  \partial^2_{\theta} F \partial^2_{v} F - 6 \csc\theta \ (\partial_v \partial_{\phi} F)^2 \nonumber\\ &-& 6 \sin\theta \ (\partial_v \partial_{\theta} F)^2 + 8 \sin\theta \ \partial_v F \partial^2_v F \Big]~.
\label{LagS}
\end{eqnarray}
\
Neglecting total derivative terms, we can write final Lagrangian as,
\begin{eqnarray}
\mathcal{L_{F}}&=& \Big[  \frac{-3}{2 (2M)^2} \csc\theta \ \partial_{\phi} F \partial_{\phi} F - \frac{3}{2 (2M)^2} \sin\theta \ \partial_{\theta} F \partial_{\theta} F \nonumber\\ &+& 4 \sin\theta \ \partial_{v} F \partial_{v} F + 4 \cos\theta \ \partial_{\theta} F \partial^2_{v} F \nonumber\\ &+& 4 \csc\theta \ \partial^2_{\phi} F \partial^2_{v} F + 4 \sin\theta \  \partial^2_{\theta} F \partial^2_{v} F\nonumber\\ &-& 6 \csc\theta \ (\partial_v \partial_{\phi} F)^2 - 6 \sin\theta \ (\partial_v \partial_{\theta} F)^2 \Big]~.\label{lags1}
\end{eqnarray}
Here the non-vanishing lower components of $N^a$ is given by 
\begin{eqnarray}	
N_r = \frac{1}{\sqrt{f(r)- (2M/r)f(r) \partial_v F +  
	2 f(r)\partial_v F +2rf(r) \partial^2_v F}}~,
\end{eqnarray} 
where $f(r)=1-2M/r$.
Hence for GHY boundary term  the action can be expressed as,
\begin{eqnarray}
\mathcal{S}_2 &=& -  \frac{M}{8 \pi G}\int d^3 x \sin\theta  \Big[1 +  (\partial_v F + 2M  \partial^2_v F)\nonumber\\ &+& (2M  \partial_v F  \partial^2_v F + 8 M^2  (\partial^2_v F)^2 + 8 M^2  \partial_v F \partial^3_v F\nonumber\\  &+& 32 M^3  \partial^2_v F \partial^3_v F) \Big ]~,
\label{K}
\end{eqnarray}
which is again  does not contribute to the equation of motion as was the case for Rindler space.
The dynamics of the Goldstone mode will be governed by the action corresponding to  $\mathcal{L_{F}}$, and the equation of motion is given by,
\begin{eqnarray}
&& -8 \sin\theta \partial^2_v F + \frac{3}{(2M)^2} \cos\theta  \partial_{\theta} F + \frac{3}{(2M)^2} \sin\theta  \partial^2_{\theta} F \nonumber\\ &+& \frac{3}{(2M)^2} \csc\theta \partial^2_{\phi} F -16 \sin\theta \partial^2_v  \partial^2_{\theta} F-16 \cos\theta \partial^2_v  \partial_{\theta} F\nonumber\\ &-& 16 \csc\theta \partial^2_v  \partial^2_{\phi} F = 0~. 
\label{eqn} 
\end{eqnarray}
In this analysis full metric has been considered. Since we are interested in the near horizon symmetries, the near horizon metric could be enough to obtain the same result. For completeness, we explicitly demonstrated this in Appendix \ref{App1}.

Since the action has the rotational symmetry, we can take the following solution ansatz for Goldstone boson mode in terms of spherical harmonic, 
\begin{equation} 
F(v,\theta,\phi)=\frac{1}{k} \sum_{lm} {c_{lm}}f_{lm}(v) Y_{lm}(\theta,\phi)\label{F1},
\end{equation}
with $c_{lm}$ are constant coefficients and $f_{lm}$ are the time dependent mode function. This is consistent with the spherically symmetric Schwarzschild geometry. The factor ${1}/{k} = {4M}$ is introduced for dimensional reason.
Substituting the form of $F$ (\ref{F1}) in (\ref{eqn})  we get  following equation of motion for $f_{lm}(v)$ 
\begin{eqnarray} \label{modeeq2}
&& [2l(l+1) -1] \partial^2_{v} f_{lm} - \frac{3}{32 M^2} l(l+1) f_{lm} = 0 .
\end{eqnarray}
Since the near horizon geometry of the Scwarzchild black hole is Rindler with sphere as spatial section, one notices some significant differences in the mode dynamics governed by eq. (\ref{modeeq2}) and that of the previous case in eq.(\ref{modeeq1}). Most importantly, for spatial spherical geometry the effective potential perceived by every individual mode parametrized by $(l,m)$ is no longer universal but dependent upon the angular momentum $l$. Before we discuss the implications of this dependence, let us take a look at the behaviour of individual modes. 
\begin{itemize}
\item For $l=0$ mode, the equation reduces to,
\begin{equation}
\partial^2_{v} f_{00}(v) =0~.
\end{equation} 
 The solution of the above equation is $f_{00} = c_1(x^A) v + c_2(x^A)$. By choosing $c_1=0$,  the final  solution will be $f_{00}(v) = c_2(x^A)$.
\item For all remaining modes $l\geq 1$, we get the inverted harmonic oscillator potential similar to our previous case. One important difference is the angular momentum dependence of the inverted harmonic potential. Therefore, the universality of all the modes with respect to their time dynamics is lost as opposed to our previous study in Rindler metric with spatial section. However, it can be checked that numerically the inverted potential depends very weakly on the value of $l$, which we will discuss in terms of temperature in the next subsection. Nonetheless, the mode equation looks likes, 
\begin{eqnarray}
\partial^2_{v} f_{lm} -k^2 \Omega^2 f_{lm}(v) = 0~,
\end{eqnarray}
where, 
\begin{equation}
\Omega = \sqrt{\frac{3 l (l+1)}{ 2(2 l(l+1)-1)}}~.
\end{equation} 
We get the inverted harmonic oscillator potential similar to our previous case.
\end{itemize} 
The complete solution for all modes can therefore be,\\
\begin{itemize}
\item for $l=0$ ;
\begin{eqnarray}
F(x^A) = \sum_{lm}  \frac{1}{k} c_2(x^A) Y_{lm}(x^A);
\end{eqnarray}
\item for $l \geq 1$,
\begin{eqnarray}
F(v,x^A) = \sum_{lm} \frac{A}{k}  e^{ k v}  Y_{lm}(x^A)~. 
\label{solSR}
\end{eqnarray}
\end{itemize}

Hereafter we can proceed along the same line as discussed before. Important difference would be the state dependent inverted harmonic potential 
\begin{equation}
V_{harmonic} = - \frac12 \Omega(l)^2 {k}^2 f_{lm}^2~. 
\end{equation}
Therefore, strictly speaking for the present case degenerate states will be only for $m$ within $(-l,l)$. However, let us point out that if we consider numerical values into consideration, the value of $\Omega$ is confined within a very narrow region
 \begin{equation}
\sqrt{\frac{3}{4}}\leq\Omega(l)\leq 1~.
 \end{equation}
Hence, one can approximately consider all the quantum states of the Goldstone boson parametrized by $(l,m)$ with $l\geq1$, are quasi-degenerate. 
Unlike the previous case for the Rindler spacetime with flat spatial section, the emission probability for the present case would be identified with Boltzmann distribution with temperature, 
\begin{eqnarray}
T_l = \frac{\hbar}{8 \pi M} \Omega(l)  , 
\end{eqnarray}
which will weakly depend upon the value of angular momentum quantum number $l$. Interestingly for $l=1$ mode the above expression came out exactly same as usual black hole temperature $T_{BH}$, given by the Hawking expression \cite{Hawking:1974rv}. However considering other modes we can define an average temperature 
\begin{equation}
T_{avg} =  \frac{\hbar}{8 \pi M}  \left(\frac{\sum_l \Omega(l)}{\sum_l 1}\right) = \frac{\hbar }{8 \pi M} \left(\sqrt{\frac34}\right)  = \frac{\sqrt{3}}{2} T_{BH}~,
\end{equation}
Here again we observed that the Goldstone modes are inherently thermal in nature. The obtained temperature is proportional to the Hawking expression for that of the Schwarzschild horizon.

From the analysis so far what we can infer is that since the origin of the Goldstone modes are associated with the breaking of symmetries of the horizon, those modes can be a potential candidate for the microscopic states of a black hole. Quantum mechanically all those states turned out to be thermal with a specific temperature. However, origin of different expressions for the temperature compared with that of the usual Hawking temperature needs to be explored in detail. Furthermore, nature of degeneracy of those Goldstone states appears to be dependent upon the spacetime background. Such as for Rindler spacetime with plane symmetric horizon, all the modes emerged as degenerate and, therefore, each mode fills the same temperature. On the other hand for Schwarzschild black hole this is not the case as the degeneracy of states has been lifted by the less symmetric spherical horizon. Nevertheless, we hope that this thermal nature of the Goldstone modes at the quantum level can be inferred for all types of horizon. We keep this for our future project.

\section{Summary and conclusions}
Microscopic origin of the thermodynamic nature of the black hole is one of fundamental questions in the theory of gravity. It is obvious that within the framework of Einsteinian gravity this question can not be answered. However, the recent understanding of infrared behavior of gravity opens up a new avenue towards understanding this question. In the gravitational theory, one of the interesting infrared properties is the emergence of infinite dimensional symmetry at null infinity which leads to soft graviton theorem. Over the years it has been observed that analogous symmetry exists near the null horizon which can play important role in explaining the microscopic origin of horizon thermodynamics. Here we particularly concentrated on the BMS-like symmetry in the near horizon region. Under the diffeomorphism symmetry, appropriate boundary conditions are imposed in such a way that the near horizon form of the metric remains unchanged. It is observed that in this process the macroscopic parameters, like mass (surface gravity), get modified. This change in macroscopic parameters is argued to be the phenomena of symmetry breaking on the horizon and corresponding parameter can be viewed as the Goldstone mode.   

In the present paper our main effort was to explore the dynamics of these Goldstone modes. For the purpose of our present study, we consider two simple gravitational background. One is simple Rindler spacetime with flat Killing horizon and the other one is Schwarzschild black hole.   
Our preliminary investigation at tree level reveals  that the horizon is indeed a special place where the dynamics of the Goldstone mode in momentum space is governed by inverse harmonic potential. As mentioned earlier, in the framework of classical Einsteinian gravity it is difficult to understand this situation as those modes are simply unstable.  Interestingly, at the quantum level this instability \cite{Morita:2019bfr} can have a nice interpretation in terms of inherent  thermality in connection with its chaotic behaviour, which may provide us a first glimpse of microscopic view of the horizon thermodynamics. Interestingly, for both the gravitational backgrounds, as expected the temperature turned out to be proportional to the surface gravity which is similar to the expression (except a numerical factor) given by Unruh \cite{Unruh:1976db} and Hawking \cite{Hawking:1974rv}. This led us to think that these Goldstone modes might be candidates for the microscopic description of the horizon thermality.
Even more interestingly, we found out  large number of degenerate states for Rindler and qusi-degenerate states for Schwarzschild black holes which may be responsible for the horizon entropy. We will take up these issues in more detail in future publication.

So far we have considered the black hole spacetime which are static and hence generating only one Goldstone field. However for a gravitational background having intrinsic rotation such as Kerr spacetime, corresponding analysis of the Goldstone mode dynamics will be more effective. This is because in this case there will be more than one symmetry generator. This topic is now under investigation. Finally, we want to mention that since the Goldstone modes are thermal in nature, it might be interesting to look at BMS symmetry in this way hoping that such an analysis will be able to shed some light towards the microscopic description of horizon thermodynamics. 


\appendix
\section{\label{App1}Near horizon analysis of Schwarzschild black hole}
As mentioned in the main text, in this section we will argue that same dynamical equations and solution for the Goldstone modes can be obtained starting from  the near horizon metric of the Schwarzschild black hole. 
Now we can easily check that the action constructed from the near horizon metric will contains three types of terms. The ones which are independent and linear in $F$, can be traced back from their origin which can be transmitted to the fact that the near horizon geometry of the Schwarzschild black hole is Rindler times a sphere, and it does not satisfy the background Einstein's equation. We, therefore, ignore those terms as they can also be made total derivative. Non-trivial dynamics of the Goldstone modes are attributed to second order term in $F$ in the action, and it can be easily checked that those terms are exactly the same as in (\ref{LagS}) up to a total derivative. As a result with a proper prescription, full spacetime geometry as well as near horizon geometry of the Schwarzschild background are giving rise to the same Goldstone mode dynamics. 

{\section{\label{App2} Surface Hamiltonian and heat content} 
In the main text we have constructed the GHY boundary term in the action formulation which did not contribute in the dynamics of the Goldstone mode. However an important analysis is left to be discussed there. It is well known that the boundary term of the Einstein-Hilbert action in gravitational theroy leads to surface Hamiltonian which is directly related to the heat content of the Horizon (detail discussion is given in (\cite{Padmanabhan:2012bs} \cite{Majhi:2013jpk})). The expression of the surface Hamiltonian comes out to be the product of temperature and entropy of the horizon. Keeping this in mind we can write surface Hamiltonian corresponding to the GHY boundary term (\ref{Kac}):
\begin{eqnarray}
H_{sur} = -\frac{\partial S_2}{\partial v}~. \label{H}
\end{eqnarray}
 Now substituting the solution (\ref{gsol2}) in the expression (\ref{Kac}) and integrating the boundary term , the Hamiltonian (\ref{H}) comes out:
 \begin{eqnarray}
H_{sur}=\frac{\bar{A}}{8 \pi G} \Big[ \alpha + f_3(\alpha) 
 e^{(f_2(\alpha) v)}\Big]~,
\end{eqnarray}
 	 $ \bar{A}$ denotes the transverse area of the Rindler horizon. In the second part function $f_3(\alpha)$  comes from the first and higher order time derivative of $F$ in the boundary term of the action. Whereas $f_2 (\alpha)$ denotes all the square, cubic and higher order terms in the expression.  Now near horizon (where $v \rightarrow -\infty$) the second terms in the above expression vanishes. Hence the result comes out as:
 \begin{eqnarray}
 H_{sur} = \frac{1}{8 \pi G} \bar{A} \alpha = TS~, \label{TS}
\end{eqnarray}  
where $S = \bar{A} /4 G $  and $T = \alpha/2\pi$ are  the Horizon entropy and temperature respectively.The result clearly indicates that irrespective of Goldstone modes $F_{m,n}$, the GHY boundary term in the action is related to the heat content of the horizon. Similar conclusion can be drawn for Schwarzschild black hole horizon also. 

\end{document}